\documentclass[12pt]{article}
\usepackage{epsf}
\usepackage{amsmath}
\usepackage{graphics}
\usepackage{cite}

\renewcommand{\thefootnote}{\#\arabic{footnote}}

\newcommand{\gtrsim}{ \mathop{}_{\textstyle \sim}^{\textstyle >} }

\begin{document}

\setcounter{footnote}{0}
\begin{titlepage}

\begin{center}

\vskip .5in

{\large \bf
On the Determination of Neutrino Masses and Dark Energy Evolution from the Cross Correlation of CMB and LSS
}

\vskip .45in

{\large
Kazuhide Ichikawa and Tomo Takahashi\footnote{
Present address: Department of Physics, Saga University, Saga 840-8502, Japan
}
}

\vskip .45in

{\em
Institute for Cosmic Ray Research,
University of Tokyo\\
Kashiwa 277-8582, Japan
}

\end{center}

\vskip .4in

\begin{abstract}
We discuss the possibilities of the simultaneous determination of the
neutrino masses and the evolution of dark energy from future
cosmological observations such as cosmic microwave background (CMB),
large scale structure (LSS) and the cross correlation between
them. Recently it has been discussed that there is a degeneracy between
the neutrino masses and the equation of state for dark energy. It is
also known that there are some degeneracies among the parameters
describing the dark energy evolution. We discuss the implications of
these on the cross correlation of CMB with LSS in some
details.  Then we consider to what extent we can determine the neutrino
masses and the dark energy evolution using the expected data from CMB,
LSS and their cross correlation.

\vspace{1cm}

\end{abstract}

\end{titlepage}

\renewcommand{\thepage}{\arabic{page}}
\setcounter{page}{1}
\renewcommand{\thefootnote}{\#\arabic{footnote}}
\renewcommand{\theequation}{\thesection.\arabic{equation}}

\section{Introduction} \label{sec:introduction}
\setcounter{equation}{0}

Probing the masses of neutrinos has become one of the important targets in
cosmology.  Although neutrino oscillation experiments have well measured
the mass differences as $\Delta m^2_{12} \simeq 7\times10^{-5}~{\rm
eV}^2$ from solar neutrino experiments
\cite{Cleveland:1998nv,Abdurashitov:2003ew,Kirsten:2003ev,Fukuda:2001nk,Aharmim:2005gt,Araki:2004mb}
and $\Delta m^2_{23} \simeq 2.6\times10^{-3}~{\rm eV}^2$ from
atmospheric neutrino experiments \cite{Fukuda:1998mi,Ashie:2005ik}, they
are insensitive to the absolute values of the neutrino masses. Experiments
using kinematical probe such as tritium decay measurements can give an
upper bound on an absolute neutrino mass, however cosmology can give
a more stringent bound. 
From the analyses of recent cosmological observations such as
cosmic microwave background (CMB), large scale structure (LSS) and so on,
we can conservatively say the current bound on the sum of
neutrino masses $\sum m_\nu$ is around 2 eV (95\% C.L.)  
\cite{Spergel:2003cb,Tegmark:2003ud,Seljak:2004xh,Ichikawa:2004zi,Spergel:2006hy,Tegmark:2006az,neutrino}.

Another important issue in cosmology is to understand the nature of dark
energy. Almost all current cosmological observations suggest that the
present universe is dominated by an enigmatic component called dark
energy.  Although many models for dark energy have been proposed so far,
we have not pinned down the model yet. However, by parameterizing the dark
energy with its equation of state $w_X$, cosmological observations can
give the constraints on $w_X$. Assuming $w_X$ being constant, the
current observations give $w_X \sim -1$
\cite{Spergel:2003cb,Tegmark:2003ud,Seljak:2004xh,Spergel:2006hy,Tegmark:2006az,Tonry:2003zg,SN,MacTavish:2005yk}. 
Furthermore, even if we allow time-varying $w_X$ and/or non-flat universe, it is shown that 
$w_X$ is constrained to be around a cosmological constant for some types of time variation 
\cite{DE_curv}.

It has been  shown that the constraint on the neutrino masses from
CMB, LSS, type Ia Supernovae (SNeIa) and so on can be weakened if the
equation of state for dark energy is allowed to take a value other than $-1$
and vice versa \cite{Hannestad:2005gj}, which means that there is a degeneracy between
the neutrino masses and the equation of state for dark energy.  In
Ref.~\cite{Hannestad:2005gj}, the equation of state for dark energy is
assumed to be constant in time, however, most models of dark energy
proposed so far have a time-varying equation of state. Importantly it
should be noticed that there are degeneracies among 
parameters which describe the time dependence
of dark energy equation of state. For example, it was discussed that,
for the models with a constant equation of state $w_X$ and a
time-varying equation of state parametrized as $w_X=w_0+(1-a)w_1$, there
is a degeneracy between $w_0$ and $w_1$ in the CMB power spectrum and matter power
spectrum \cite{Pogosian:2005ez}.  Since now we know that neutrino has a
mass and the equation of state for dark energy can have the time
dependence, we should take both of them into account when we consider
the constraints from observations.  However, considering the
degeneracies discussed above, we can expect that we encounter
unfortunate situations when we want to determine the neutrino masses and
the evolution of dark energy simultaneously.

In this paper, we discuss to what extent we can determine the neutrino
masses and the evolution of dark energy, i.e., the time dependent
equation of state, simultaneously from future cosmological
observations. For this purpose, we make use of the Fisher matrix
analysis using the expected data from future CMB and LSS observations.
In addition to them, we also consider the cross correlation between CMB
and LSS, which has been attracting attention of the community in
recent years. 
In particular, we investigate the effect of the neutrino masses on the 
cross correlation, which has not been attempted previously in the literature.
 As is well known, after the universe has been dominated
by dark energy, the gravitational potential decays, which drives the
so-called late integrated Sachs-Wolfe (ISW) effect.  Since the time-evolving
gravitational potential that drives the ISW effect may also affect large
scale structure formation, the temperature fluctuation from the ISW
effect and the distribution of galaxy are considered to be correlated
\cite{Crittenden:1995ak,Peiris:2000kb,Cooray:2001ab}.  Since there is no
ISW effect in the universe dominated by matter, the detections of the
cross correlation of the ISW effect with galaxy survey is a piece of
physical evidence for the existence of dark energy. The detection of the
cross correlation between the WMAP temperature fluctuation and several
galaxy surveys have been reported in
Refs.~\cite{Fosalba:2003iy,Boughn:nature427,Boughn:2004zm,Fosalba:NMRAS,Fosalba:2003ge,Scranton:2003in,
Afshordi:2003xu,Padmanabhan:2004fy,Cabre:2006qm,Giannantonio:2006du,McEwen:2006my}.  
Many authors have discussed the cross correlation of
CMB with LSS to investigate the properties of dark energy and some other
issues
\cite{Pogosian:2005ez,Garriga:2003nm,Pogosian:2004wa,Hu:2004yd,Corasaniti:2005pq,Gaztanaga:2004sk,Cooray:2005px,Afshordi:2004kz}.
In particular, it was shown that the cross correlation of CMB with
galaxy survey can be a good probe for the evolution of dark energy
\cite{Pogosian:2005ez}.

The organization of this paper is as follows. In the next section, first
we briefly discuss the degeneracy between the neutrino masses and the
equation of state for dark energy and also that in models with the
time-varying equation of state. Then, in section \ref{sec:crosscorrelation}, we discuss the cross
correlation of CMB with LSS and its spectra in scenarios with
massive neutrinos and dark energy with some equations of state.  In
section \ref{sec:perturbation}, we study effects of massive neutrinos and dark energy on the
suppression of growth of perturbation and the ISW effect and discuss how
the masses of neutrinos and the equation of state for dark energy can
affect the cross correlation spectrum.  In section \ref{sec:future}, we discuss a future
constraint on the neutrino masses and the evolution of dark energy from
CMB, LSS and the cross correlation of CMB and LSS. We give the
summary of this paper in the final section.

\section{Degeneracies in neutrino and dark energy sectors} \label{sec:degeneracy}

In this section, we briefly discuss the degeneracy between the neutrino
masses and the equation of state for dark energy $w_X$, which was
recently pointed out \cite{Hannestad:2005gj} and also the degeneracy in
the parameters which describe the time dependent $w_X$.

\begin{figure}[t]
\begin{center}
\scalebox{1}{\includegraphics{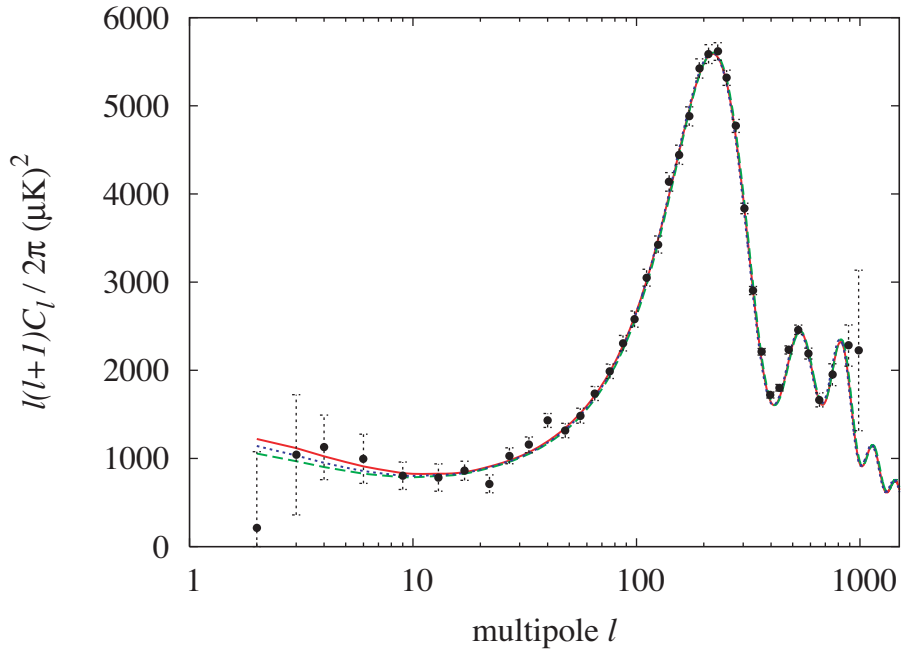}} 
\caption{ The CMB TT power spectrum for the $\Lambda$CDM model (red solid
line), the model with $w_X = -1.5$ and $\Omega_\nu h^2= 0.01$
(green dashed line) and the model with the time-varying equation of
state $w_0=-1 - 1.2(1-a)$ and $\Omega_\nu h^2= 0.01$ 
(blue dotted line). Notice that three lines are almost
 indistinguishable. The data from WMAP \cite{Spergel:2006hy} are
also plotted. For the cosmological parameters taken here, see the
text. }
\label{fig:cmb}
\end{center}
\end{figure}

\begin{figure}[t]
\begin{center}
\scalebox{1}{\includegraphics{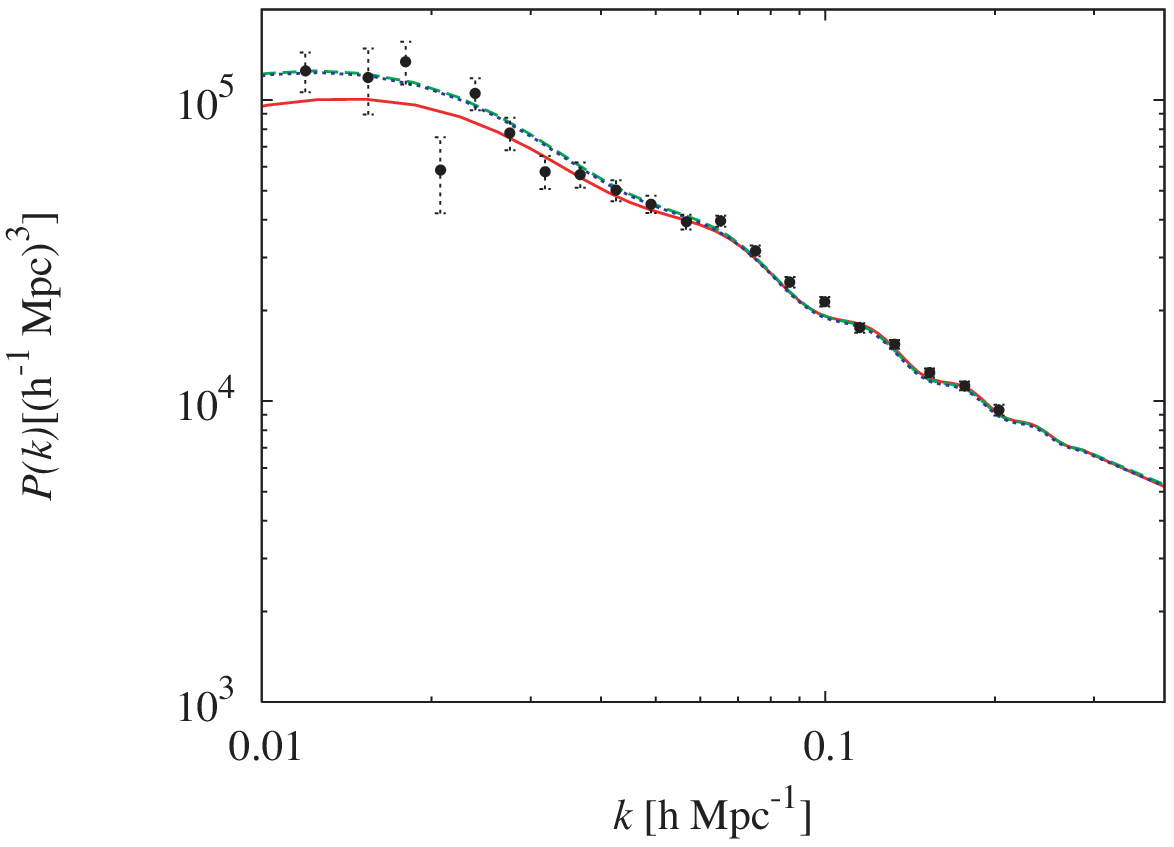}} 
\caption{Matter power spectrum for the same models given in
Fig.~\ref{fig:cmb}. The data from SDSS \cite{Tegmark:2006az} are also plotted.}
\label{fig:Pk}
\end{center}
\end{figure}

First we discuss the degeneracy between the neutrino masses and $w_X$.
In Figs.~\ref{fig:cmb} and \ref{fig:Pk}, we show the CMB power spectra
and matter power spectra respectively, for the cases with $\Lambda$CDM
model and a model with massive neutrinos and the dark energy equation of
state $w_X$ which is not equal to $-1$.  For reference, we also plot the
data from WMAP3 \cite{Spergel:2006hy} in Fig.~\ref{fig:cmb} and those from SDSS \cite{Tegmark:2006az} in
Fig.~\ref{fig:Pk}.  For the $\Lambda$CDM model, we take the cosmological
parameters as $\Omega_mh^2=0.13$, $\Omega_bh^2=0.023$, $h=0.735$, $\tau=0.09$ and $n_s=0.95$ which are the mean value of 
the power-law $\Lambda$CDM models from WMAP3.  Here $\Omega_i$ is the
energy density for a component $i$ ($i$ can be $b$, $m$ and $\nu$, which respectively stand for baryon, matter and neutrino)
normalized by the critical energy
density, $h$ is the Hubble parameter, $\tau$ is the reionization optical
depth and $n_s$ is the scalar spectral index of primordial
fluctuation. In this paper, we assume that the tensor mode is
negligible.
For the matter power spectra, we have corrected for a scale-dependent biasing
following the treatment of Ref.~\cite{Tegmark:2006az}. This correction affects only the scales
with $k \gtrsim 0.07\,h\,$Mpc$^{-1}$.
 For the model with massive neutrinos and dark energy with
$w_X \ne -1$, we take the cosmological parameters as $\Omega_\nu h^2=0.01$ and $w_X=-1.5$.  Here we assume degenerate masses for 3 neutrino
flavors and a constant $w_X$.  The energy density and the masses of
neutrinos are related by $\Omega_\nu h^2 = \sum m_\nu / 93.2\,{\rm
eV}$.  The other cosmological parameters are taken as $\Omega_m h^2 =
0.13$, $\Omega_b h^2 = 0.023$, $h=0.74$, $\tau=0.087$ and $n_s=0.94$.  In
general, dark energy component is characterized with its equation of
state $w_X $ and speed of sound $c_s^2 =\delta p_X /\delta
\rho_X$. Although the background evolution depends on $w_X$ only, the
fluctuation of dark energy depends on both $w_X$ and $c_s^2$.  In this
paper, we assume that $c_s^2=1$, which corresponds to the case with a
scalar field dark energy, where $c_s^2$ is defined at the rest frame of
the dark energy.  As seen from the figure, these two parameter sets give
almost the same spectra, which means that the values of $\chi^2$ from
CMB and LSS are almost the same.
 As shown in
Ref.~\cite{Hannestad:2005gj}, there is a degeneracy between $m_\nu$ and
$w_X$ in the direction to the region where $w_X < -1$. This degeneracy
can be explained as follows.  To fit a model with massive neutrinos to
large scale structure data, large $\Omega_\nu h^2$ can be compensated by
increasing the energy density of matter. This cancellation seems not to
be allowed because, assuming a flat universe and the cosmological
constant for dark energy, SNeIa data constrain $\Omega_m$ to be $\sim
0.3$.  However allowing the equation of state $w_X$ to vary, large
values of $\Omega_m$ are consistent with observations of SNeIa with
smaller $w_X$ \cite{Tonry:2003zg}.  This gives rise to the degeneracy in
the matter power spectrum.  As for CMB data, it is known that massive
neutrinos affect the structure of acoustic peaks such as the position
and the shape \cite{Ichikawa:2004zi}.  Increasing $\Omega_\nu h^2$, the
peak position shifts to lower multipoles. However, by decreasing $w_X$,
the position of the peak shifts to higher multipoles, which cancels the
shift caused by increasing the value of $\Omega_\nu h^2$. Although the
shape of the peaks is also slightly modified by increasing $\Omega_\nu
h^2$, by changing other cosmological parameters, we can fit such a model
to CMB data well. Thus by varying $w_X$, larger values of $\Omega_\nu
h^2$ become to be allowed by the data.

Next we discuss the degeneracy in models with dark energy with
a time-varying equation of state. As mentioned above, most models of dark
energy proposed so far such as quintessence have a time-dependent equation
of state. Thus when we consider the constraints on the equation of
state for dark energy $w_X$ phenomenologically, we should accommodate
the time evolution of $w_X$ in some way. Although several ways have been
proposed to include the time dependence of $w_X$, here we assume a
simple form as \cite{Linder:2002et,Caldwell:2003vp,Seljak:2004xh}
\begin{equation} 
w_X (a) = w_0 + (1-a)w_1.
\end{equation}

In Figs.~\ref{fig:cmb} and \ref{fig:Pk}, the CMB and matter power
spectra for the cases with $w_0=-1$ and $w_1=-1.2$ are also
plotted. Other cosmological parameters are taken to be the same as 
the case of a constant equation of state with massive neutrinos just above.
 As seen from the figure, the
degeneracy in the CMB and LSS exists between the models with a constant
$w_X$ and a time-varying $w_X$. The degeneracy comes from the fact that
both models give almost the same angular diameter distance from  last
scattering surface which determines the positions of the acoustic peaks
in the CMB angular power spectrum.  Furthermore, the fluctuation of dark
energy is almost irrelevant to the structure of the acoustic peaks\,\footnote{For the parameters assumed here, dark energy is negligible at the epoch of recombination.},
hence they give almost an identical power spectrum provided the angular
diameter distances from last scattering surface are the same. A possible
difference between the models with constant and time dependent $w_X$ is
the ISW effect which affects the low multipole region. In fact, there is
a small difference in the region, however, it is not enough to
differentiate them due to the cosmic variance in this case.

Since there is a degeneracy between the neutrino masses and the constant
equation of state for dark energy, we can expect that the situation
becomes much worse when we also consider the time dependent dark energy
equation of state. However, the possibility of differentiating the
models with constant and time dependent $w_X$ using the cross
correlation of CMB with LSS has been discussed
\cite{Pogosian:2005ez}. Thus we may be able to break these degeneracies
using the cross correlation. Before we discuss the future constraints 
on the masses of neutrino and dark energy evolution
from CMB, LSS and their cross correlation, we study the effects of them
on the cross correlation of CMB with LSS in
the next section.

\section{Cross correlation of CMB with galaxy} \label{sec:crosscorrelation}

In this section, first we briefly review the formulation of the
cross correlation of CMB with galaxies following
Ref.~\cite{Garriga:2003nm}. Detailed descriptions of this issue can be
found in Refs.~\cite{Peiris:2000kb,Cooray:2001ab,Garriga:2003nm}.

The two point correlation function between the ISW signal and
galaxy overdensity is defined as
\begin{equation}
C^{\rm ISW-g} (\theta) 
=
 \left\langle \Delta_{\rm ISW} (\hat{n}_1) 
\delta_{\rm g} (\hat{n}_2)\right\rangle,
\end{equation}
where the angular brackets represent the ensemble average and $\cos
\theta = \hat{n}_1 \cdot \hat{n}_2$. $\Delta_{\rm ISW}$ is the
temperature fluctuation from the ISW effect in the direction $\hat{n}_1$
which is written as
\begin{equation}
\Delta_{\rm ISW} (\hat{n}_1) 
= \frac{\Delta T_{\rm ISW}}{T} 
= -2 \int e^{-\tau(\eta)} 
\Phi' (\hat{n}_1, \eta)d\eta,
\end{equation}
where $\eta$ is the conformal time and $e^{-\tau(\eta)}$ is the
visibility function. Here a prime denotes the derivative with respect to
the conformal time. $\Phi$ is the gravitational potential which appears
in the metric perturbation in the conformal Newtonian gauge as
\begin{equation}
 ds^2 = - a^2 ( 1 + 2 \Phi)d\eta^2 + a^2 (1 + 2 \Psi) dx^2.
\end{equation}
Unless otherwise stated, we work in the conformal Newtonian gauge in
this paper.

$\delta_{\rm g} (\hat{n}_2)$ is the overdensity of galaxies
in the direction $\hat{n}_2$.
\begin{equation}
\delta_{\rm g} (\vec{n}_2)
= \frac{n_{\rm g} (\vec{n}_2)- \bar{n}_{\rm g}}{\bar{n}_{\rm g}},
\end{equation}
where $\bar{n}_{\rm g}$ is the mean number density of galaxies.  It is
assumed that the galaxy number overdensity  traces the density
fluctuation of matter as
\begin{equation}
\delta_{\rm g} = b \delta_{\rm m},
\end{equation}
where $b$ represents the galaxy bias.
Thus $\delta_{\rm g}$ can be written as 
\begin{equation}
\delta_{\rm g} (\hat{n}_2) 
= b \int \phi(z) z' \delta_m (\hat{n}_2, \eta) d\eta,
\end{equation}
where $\phi(z)$ is the selection function of a given galaxy survey.

As usual, it is convenient to decompose $C^{\rm ISW-g} (\theta)$
with the Legendre polynomials $P_l(\cos \theta)$ as
\begin{equation}
C^{\rm ISW-g} (\theta) 
=
\sum_{l=2}^{\infty}  
\frac{2l+1}{4 \pi} C_l^{\rm ISW-g} P_l (\cos \theta).
\label{eq:Ctheta}
\end{equation}
We subtracted the monopole and dipole contributions by construction.
With the above definition, the cross-correlation power spectrum is
given by
\begin{equation}
C^{\rm ISW-g}_l
= 
4 \pi \frac{9}{25} \int \frac{dk}{k} 
\Delta_\mathcal{R}^2 I_l^{\rm ISW} (k) I_l^{\rm g}(k),
\end{equation}
where $\Delta^2_\mathcal{R}$ is the primordial power spectrum.
$I_l^{\rm ISW} (k)$ and  $I_l^{\rm g}(k)$ are defined as
\begin{eqnarray}
I_l^{\rm ISW} (k) 
&=& 
 -2 \int e^{-\tau(\eta)} \Phi'_{k}  (\eta) j_l [ k(\eta_0 - \eta)] d\eta, \\
I_l^{\rm g} (k) 
&=&
 b \int \phi(z) z' \delta_m^k (\eta) j_l [ k(\eta_0 - \eta)] d\eta,
\end{eqnarray}
where the $j_l$ is the spherical Bessel function and $\eta_0$ is the
conformal time at the present epoch.  $\Phi_k$ and $\delta_m^k$ are the
Fourier component of the gravitational potential and density 
fluctuation of matter which can be written as 
\begin{eqnarray}
\Phi ( r \hat{n} ) 
&=&
\int \frac{d^3 k}{(2\pi)^3} \Phi_k(\hat{n}) e^{i \vec{k}\cdot \hat{n} r }, \\
\delta_m ( r\hat{n} ) 
&=&
\int \frac{d^3 k}{(2\pi)^3} \delta_m^k(\hat{n}) e^{i \vec{k}\cdot \hat{n} r}, 
\end{eqnarray}
where $r \equiv \eta_0 - \eta$. For the later use, we also write down
here  the galaxy-galaxy correlation function which is given as 
\begin{equation}
\omega_l^{(i,j)} = 4 \pi \frac{9}{25} \int \frac{dk}{k} 
\Delta_\mathcal{R}^2 I_l^{\rm g(i)} (k) I_l^{\rm g(j)}(k),
\end{equation}
where indices $(i,j)$ represent the redshift bins for the galaxy
selection functions.

\begin{figure}[t]
\begin{center}
\scalebox{1}{\includegraphics{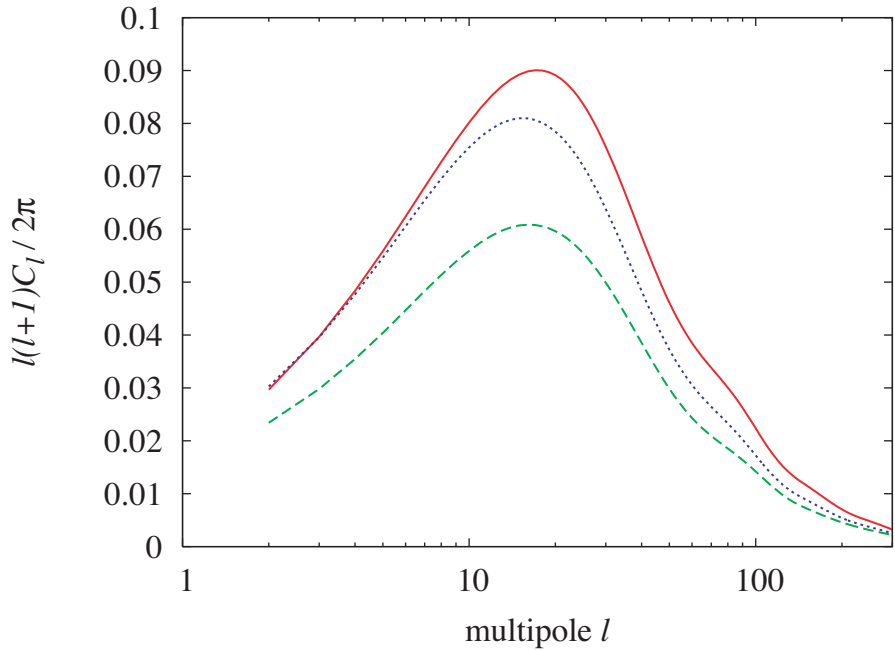}} 
\caption{ The cross correlation power spectrum for the same models given
 in Fig.~\ref{fig:cmb}. }
\label{fig:cross}
\end{center}
\end{figure}

Now we show the cross correlation spectrum in models introduced in the
previous section.  We computed the cross correlation spectra using the
modified version of CMBFAST \cite{Seljak:1996is}.  In
Fig.~\ref{fig:cross}, we plot the cross correlation power spectrum for
the same models as those in Fig.~\ref{fig:cmb}.  Here we use the
normalized Gaussian selection function with the peak at redshift
$\bar{z}=0.5$ and the variance $\sigma_z=0.07$.  The other cosmological
parameters are the same as those in Fig.~\ref{fig:cmb}.  In the figure,
the scalar amplitudes are normalized to the WMAP best fit values and the
bias factors are determined by the fit to SDSS data.  
We will investigate in Sec.~\ref{sec:future} whether or not the difference seen in Fig.~\ref{fig:cross} can be probed by future cross-correlation data.
Before that, in the next section, we investigate the cross correlation in detail,
discussing the suppression of matter fluctuation and the ISW effect in
models with massive neutrinos and dark energy with some equations of
state.

\section{Suppression of perturbation growth and the ISW effect} \label{sec:perturbation}

In this section, we discuss the effects of massive neutrinos and dark
energy with some equations of state on the suppression of fluctuation
growth and the ISW effect. 
Below, the evolution of perturbation variables are calculated
at the mode $k=0.01h$\,Mpc$^{-1}$.
Then we study the cross correlation spectrum
in the models.

\begin{figure}[t]
\begin{center}
\scalebox{1}{\includegraphics{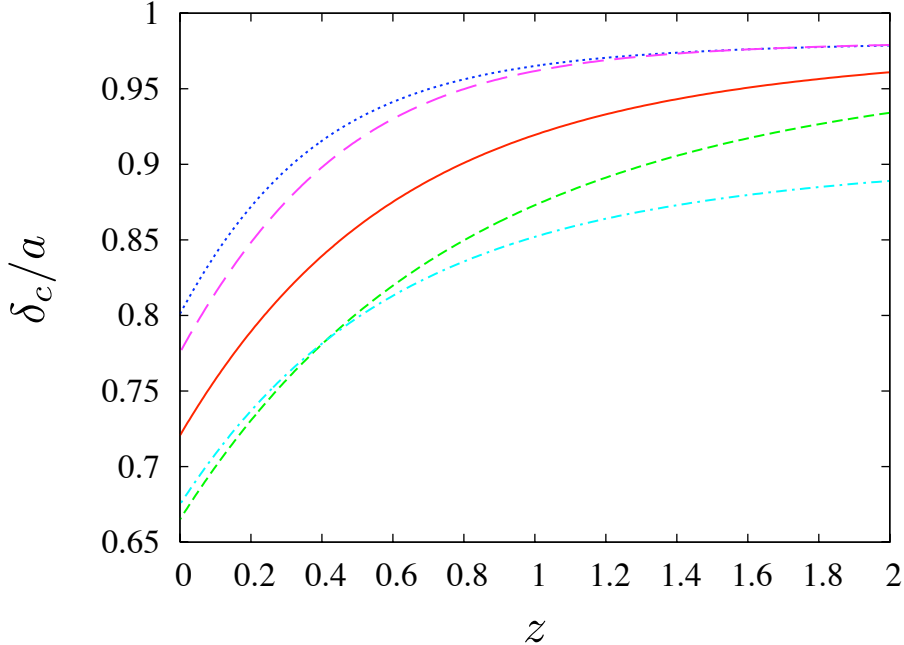}} \caption{Suppression of
 fluctuation growth in models with dark energy equations of state $w_X=-1$
 (red solid line), $w_X=-0.8$ (green short-dashed line), $w_X=-1.5$
 (blue dotted line) and $w_X=-1-1.2(1-a)$ (purple long-dashed line).
In these cases, we assume that $\Omega_\nu h^2 =0$.
 The case with massive neutrinos $\Omega_\nu h^2=0.01$ and $w_X=-1$ (light blue dash-dotted line) is also plotted.}
\label{fig:delta}
\end{center}
\end{figure}

\begin{figure}[t]
\begin{center}
\scalebox{1}{\includegraphics{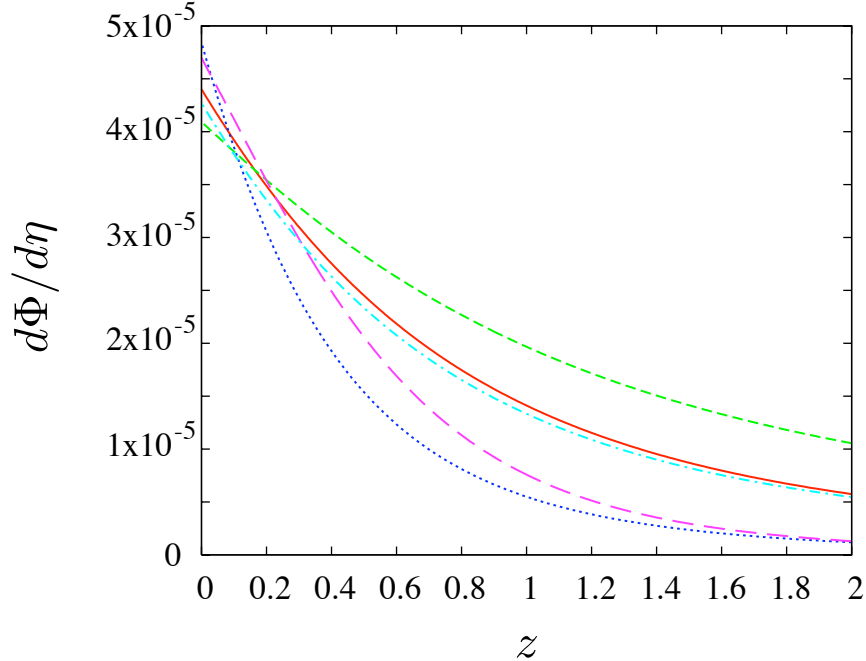}}
\caption{The derivative of the gravitational potential with respect to the 
conformal time for the same models given in Fig.~\ref{fig:delta}. }
\label{fig:phidot}
\end{center}
\end{figure}

First we consider the effects of massive neutrinos on the suppression of
fluctuation growth. Since massive neutrinos represent a smooth
gravitationally stable component on small scales, they suppress the
growth of matter fluctuation \cite{Hu:1997vi}.  In linear regime, a
small initial fluctuation $\delta_0$ grows as $\delta (a) = D(a)
\delta_0$ where $D(a)$ is the growth factor.  During matter-dominated
universe, $D(a)$ grows as $D(a) \propto a$.  An analytic formula for the
growth factor is given in Ref.~\cite{Hu:1997vi}.  In
Fig.~\ref{fig:delta}, we show $\delta_m /a$ for the model with
$\Omega_\nu h^2=0.01$ and $w_X=-1$ as a function of redshift.  Other
cosmological parameters are taken to be the same as those of the
$\Lambda$CDM model in Fig.~\ref{fig:cmb}.  We can see the suppression of
growth of matter fluctuation in models with massive neutrinos.  The
growth of matter fluctuation is more suppressed by increasing the masses of
neutrinos. To see the ISW effect in the model, we also plot $d\Phi /
d\eta$ as a function of redshift in Fig.~\ref{fig:phidot}.  As is clear
from the figure, massive neutrinos have little effect on the evolution
of the gravitational potential.  Thus the mass of neutrinos is almost
irrelevant to the late ISW effect, which means that we can expect that
the neutrino masses have almost no effect on the cross correlation of
CMB with LSS except for the overall suppression of matter
fluctuation. Notice that, since we do not have much knowledge of the
bias factor, we might not see such suppression.

Next we discuss the effects of dark energy on the suppression of
fluctuation growth and the ISW effect.  First we will discuss the
suppression of fluctuation growth.  The growth of matter fluctuation can
be described by the following equation \cite{Bean:2003fb}, including the
fluctuation of dark energy
\begin{equation}
\delta_c{''} + \mathcal{H}  \delta_c'  
- \frac{3\mathcal{H}^2}{2} \Omega_c  \delta_c
= \frac{3\mathcal{H}^2 }{2} \Omega_X \delta_X,
\end{equation}
where $\mathcal{H}=a'/a$ is the conformal Hubble parameter, $\delta_X$
is the density perturbation of dark energy and $\Omega_i (z)= \rho_i
(z)/\rho_{\rm total} (z)$ is the energy density of a component $i$
normalized by the total energy density at redshift $z$.

 In Fig.~\ref{fig:delta}, the fluctuations of matter for the cases with
$w_X=-1.5,-1,-0.8$ and $w_X=-1 -1.2(1-a)$ are shown.  The matter fluctuation is
more suppressed as increasing the value of $w_X$.  This is due to the
change of the background evolution by the domination of dark energy. The
fluctuation of dark energy also affects the fluctuation of matter via
gravity.

Another important effect of dark energy is the late ISW effect.  As is
well-known, when the universe is dominated by dark energy, the
gravitational potential decays, which drives the late time ISW effect.
The ISW effect can be written as
\begin{equation}
\Phi' \sim 
\frac{1}{2 k^2} 
\left[ \mathcal{H}^2 ( \Omega_m \delta_m + \Omega_X \delta_X ) \right].
\end{equation} 
In Fig.~\ref{fig:phidot}, the evolutions of $\Phi' $ as a function of
redshift for several values of $w_X$ are shown.  Increasing the values
of $w_X$, the ISW effect is more enhanced at high redshift region, which
leads to enhanced cross correlation power spectra\footnote{
In fact, if we consider a smooth dark energy, the tendency is different.
When we assume a smooth dark energy, the effect of dark energy on the
ISW effect comes from the modification of the background evolution
alone.  Thus in this case, increasing $w_X$ makes the epoch when the
universe begins to accelerate later, then the ISW effect becomes
less.
}.  Notice that, at recent epoch, the values of $\Phi'$ decreases as the
values of $w_X$ increases. Thus the effects of $w_X$ can depend on the
values of $w_X$ itself and also on the redshift.

\begin{figure}[t]
\begin{center}
\scalebox{1}{\includegraphics{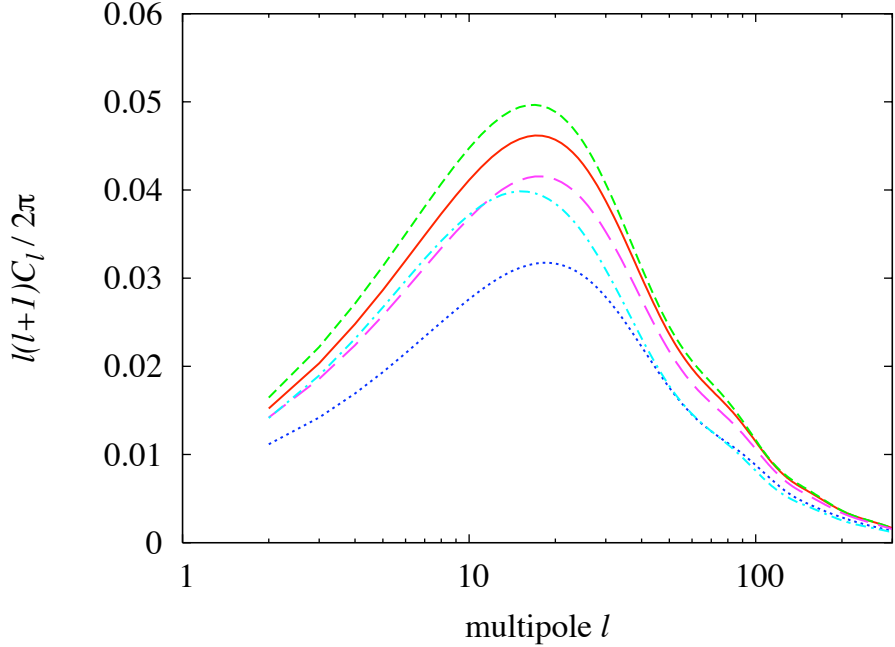}}
\caption{Cross-correlation power spectra for the same models given in
Fig.~\ref{fig:delta}.  }
\label{fig:cross2}
\end{center}
\end{figure}

Now we are in a position to discuss the cross correlation power spectrum
for models with massive neutrinos and dark energy with some equations of
state. In Fig.~\ref{fig:cross2}, we show the cross correlation power
spectra for the cases with the constant equations of state ($w_X=-0.8,
-1$ and $-1.5$) and the time-varying equation of state
$w_X=-1-1.2(1-a)$. The case with $\Omega_\nu h^2=0.01$ and $w_X=-1$ is
also plotted.  We use the same galaxy selection function as that used in
Fig.~\ref{fig:cross}.  Other cosmological parameters are taken to be the
same as those of the $\Lambda$CDM model in Fig.~\ref{fig:cmb}.  Since
the masses of neutrinos suppress the growth of fluctuation, we can
see the power spectrum of the cross correlation is also suppressed.
Although the growth of matter fluctuation can also be suppressed in the
dark energy dominated universe, effects of dark energy also come from
the ISW effect. As the equation of state of the dark energy increases,
the epoch where the energy density of dark energy dominates the universe
becomes earlier than the case with the cosmological constant, which
affects the evolution of the gravitational potential more at high
redshift. Thus it can produce the large ISW signals.  However, as mentioned
above, the size of the ISW effect also depends on the epoch we
consider. Thus the size of the cross correlation power spectrum depends
on the values of $w_X$, its time dependence (i.e., $w_1$ in our
parameterization) but also on the selection function.

\section{Future constraints} \label{sec:future}

Now we discuss attainable constraints on neutrino masses and the
evolution of dark energy from future experiments of CMB, LSS and the
cross correlation between them.  We adopt the usual Fisher matrix method
\cite{Tegmark:1996bz} for this purpose.  The Fisher matrix is defined as
\begin{equation} 
F_{\alpha \beta} = -
\left\langle \frac{ \partial^2 \ln \mathcal{L} }{\partial \lambda_\alpha
\partial \lambda_\beta} \right\rangle,
\end{equation} 
where $\mathcal{L}$ is the probability of observing a data set $\{ x_1,
x_2, ...\}$ for a given cosmological parameter set $\{\lambda_1,
\lambda_2, ....\}$.  Since there have been many works on the Fisher
matrix analysis using future CMB and LSS data, we refer
Refs.~\cite{Jungman:1995av,Tegmark:1997rp,Eisenstein:1998hr} for the
details.

As for the cross correlation, here we briefly summarize the formulation
following Ref.~\cite{Pogosian:2005ez}.  To forecast the errors in
cosmological parameters, we use the expected CMB data from PLANCK
\cite{planck} and the expected galaxy survey by Large Synoptic Survey
Telescope (LSST) \cite{LSST}. For PLANCK, we use the experimental
specification which can be found in Ref.~\cite{Lesgourgues:2004ps}. For
the distribution of galaxies from a given survey, we assume that the
total galaxy number is given by \cite{Hu:2004yd,Pogosian:2005ez},
\begin{equation} 
n_g^{\rm total}(z) \propto z^2 e^{- (z/z_n)^2},
\end{equation} 
where $z_n$ is a parameters which gives a median redshift of the
survey. The galaxies can be subdivided into multiple bins as
\begin{equation} 
n^{\rm total}_g (z)= \sum_i n_g^i (z),
\end{equation} 
through photometric redshift. To approximate the
redshift binning, we assume that photometric redshift estimates are
Gaussian distributed with the variance of $\sigma(z) =
\sigma_{\rm max} (1+z)/(1+z_{\rm max})$ \cite{Pogosian:2005ez}. Then
the photometric redshift distributions are given by
\begin{equation}
n^i_g (z) = \frac{1}{2} n_g^{\rm tot}(z) \left[ 
{\rm erfc} \left( \frac{ z_{i-1} - z}{\sqrt{2} \sigma(z)} \right)
-  {\rm erfc} \left( \frac{ z_{i} - z}{\sqrt{2} \sigma(z)} \right)
\right],
\end{equation}
where erfc is the complementary error function. For the LSST survey, we
assume the galaxy survey sky coverage fraction as $f_{\rm sky}=0.5$ and
10 photometric redshift bins out to $z \sim 3$. The total galaxy number density
is assumed  to be $70$ gal/arcmin$^2$.

For the CMB-galaxy cross correlation, the Fisher matrix can be written as
\cite{Pogosian:2005ez}
\begin{equation}
F_{\alpha \beta}^{\rm ISW-g} 
= f_{\rm sky} \sum_{l}  (2l +1) \sum_{ij} 
\frac{\partial C^{\rm ISW-g} _{l(i)} }{\partial \lambda_\alpha}  
({\rm Cov}_l)^{-1(i,j)}_{\rm ISW-g} 
\frac{\partial C^{\rm ISW-g}_{l(j)} }{\partial \lambda_\beta},
\end{equation}
where the covariance matrix $({\rm Cov}_l)_{\rm ISW-g}$ is given by 
\begin{equation}
({\rm Cov}_l)^{(i,j)}_{\rm ISW-g} 
= 
\tilde{C}^{TT}_l  \tilde{\omega}^{(i,j)}_l 
+ C^{\rm ISW-g}_{l(i)}C^{\rm ISW-g}_{l(j)}.
\end{equation}
Here $\tilde{C}^{TT}_l$ is the observed spectrum which includes the noise
\begin{equation}
\tilde{C}^{TT}_l = C^{TT}_l + \sum_C  
\left[ \left(\frac{\sigma_{C} \theta_{{\rm FWHM}, C}}{T_{\rm CMB}} \right)^{-1}
e^{l(l+1) \theta_{{\rm FWHM}, C} / 8\ln 2 }
\right]^{-1},
\end{equation}
where $\sum_C$ represents the summation over channels.  $\sigma_C$ and
$\theta_{\rm FWHM,C}$ are the sensitivity per pixel and the width of the
beam of a channel for a given measurement, respectively. $T_{\rm CMB}$
is the temperature of CMB.  $\tilde{\omega}^{(i,j)}_l$ is the observed
auto-correlation which is the sum of the signal and the Poisson noise
\begin{equation}
\tilde{\omega}^{(i,j)}_l =\omega^{(i,j)}_l + N_l^{(i,j)},
\end{equation}
where $N_l^{(i,j)}$ is the Poisson noise which is uncorrelated between bins, thus 
\begin{equation}
N_l^{(i,j)} = \frac{\delta_{ij}}{\bar{n}_A^i}.
\end{equation}
$\bar{n}_A^i$ is the galaxy number per solid angle in the $i$-th bin.
For the future constraints from CMB and LSS, we followed the
method presented in Ref.~\cite{Lesgourgues:2004ps}.

\begin{figure}[t]
\begin{center}
\begin{tabular}{cc}
\scalebox{0.7}{\includegraphics{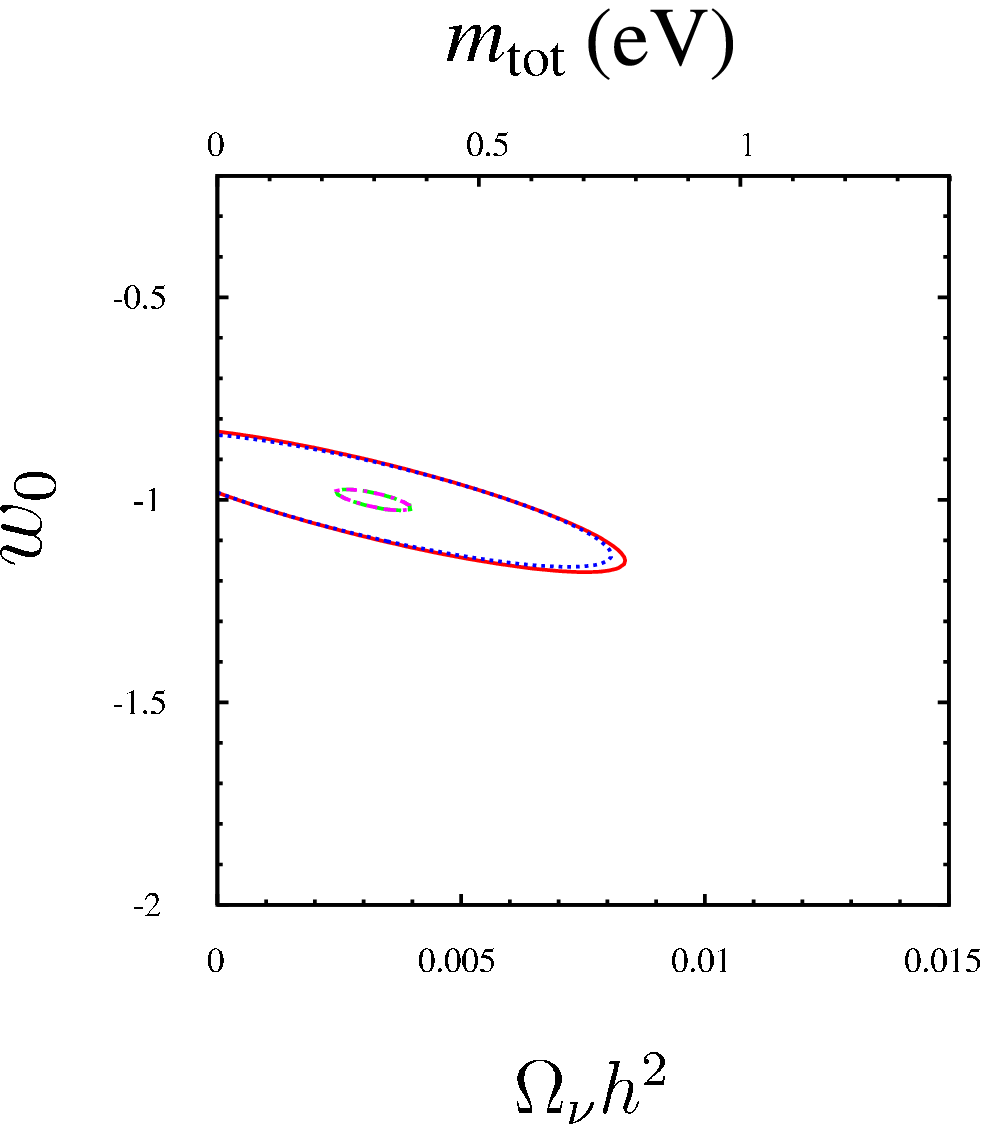}} &\hspace{10mm}
\scalebox{0.7}{\includegraphics{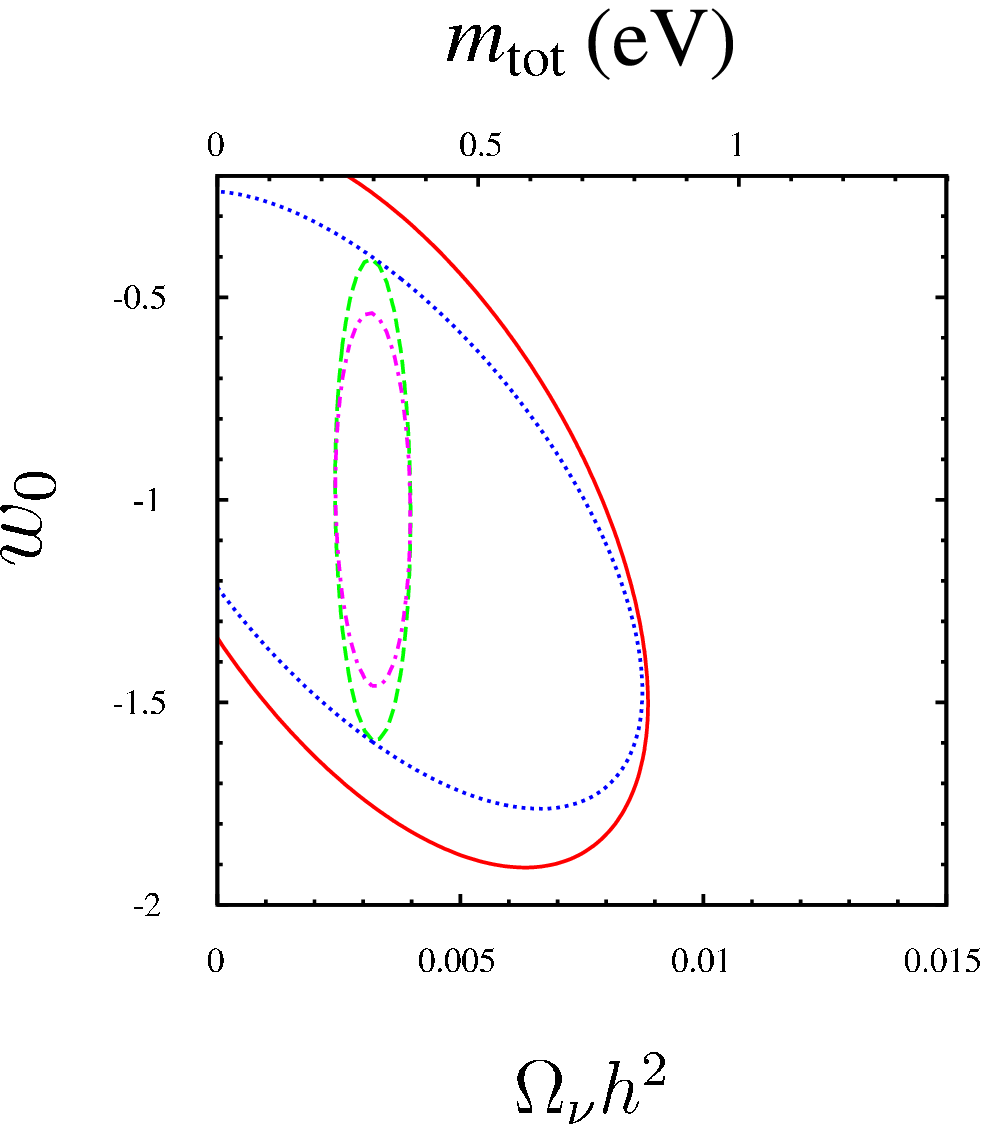}}
\end{tabular}
\caption{Expected 1$\sigma$ contours on the $\Omega_\nu h^2$ vs. $w_0$
 plane.  The cases with the constant $w_X$ (left) and time-varying $w_X$
 (right) are shown. We marginalize $w_1$ for the latter case. Contours
 which correspond to the attainable future constraints from CMB only
 (red solid line), CMB$+$cross correlation (blue dotted line), CMB$+$LSS
 (green dashed line) and all combined (purple dash-dotted line) are
 shown.}
\label{fig:m_nu_w0}
\end{center}
\end{figure}

Now we discuss attainable constraints from future experiments.  As the
fiducial model, we take the cosmological parameters as $\Omega_bh^2 =
0.024, \Omega_mh^2 =0.14, \Omega_X =0.73, \tau=0.166$ and $n_s=0.99$.
For the neutrino masses, we take $\sum m_\nu=0.3$ eV with the three
degenerate masses. We checked that even if we assume the normal
hierarchy with smaller neutrino masses, our conclusion does not change
much.  For the dark energy equation of state, we take $w_X=-1$ for the
constant case and $w_X=-1 + 0.3(1-a)$ for the time dependent case.  In
Fig.~\ref{fig:m_nu_w0}, we plot the expected constraints on the neutrino
masses and dark energy equation of state for the cases with the constant
and the time dependent $w_X$ on the left and right panels respectively.
When we obtain the constraints on the $\Omega_\nu h^2$ vs. $w_0$ plane,
we marginalized over other cosmological parameters. Especially, for the
latter case, we also marginalized over $w_1$ to obtain the constraint.
In the figure, 1$\sigma$ C.L. contours are shown for the future data
from CMB only, CMB$+$cross correlation, CMB$+$LSS, and all combined.
As we can see, for the case with constant $w_X$, the cross correlation
observations do not help much to constrain the neutrino masses and
dark energy equation of state. However, for the case with the
time-evolving dark energy equation of state, the cross correlation can
be useful to determine the time evolution of $w_X$. As for the neutrino
masses, the constraint on $m_\nu$ becomes weaker just a little in the
case with the time evolving $w_X$. In both cases, the inclusion of LSS
data can help to constrain the neutrino masses. As mentioned above, even
if we assume the normal hierarchy for the neutrino masses, the shape and
the size of the contours scarcely change.  In fact, even if we
marginalize over $w_1$, the constraint on the neutrino masses is not so
weakened. This means that the time dependence of $w_X$ (i.e., $w_1$ in
our parameterization) and $m_\nu$ are not strongly correlated.

\begin{figure}[t]
\begin{center}
\begin{tabular}{cc}
\scalebox{0.7}{\includegraphics{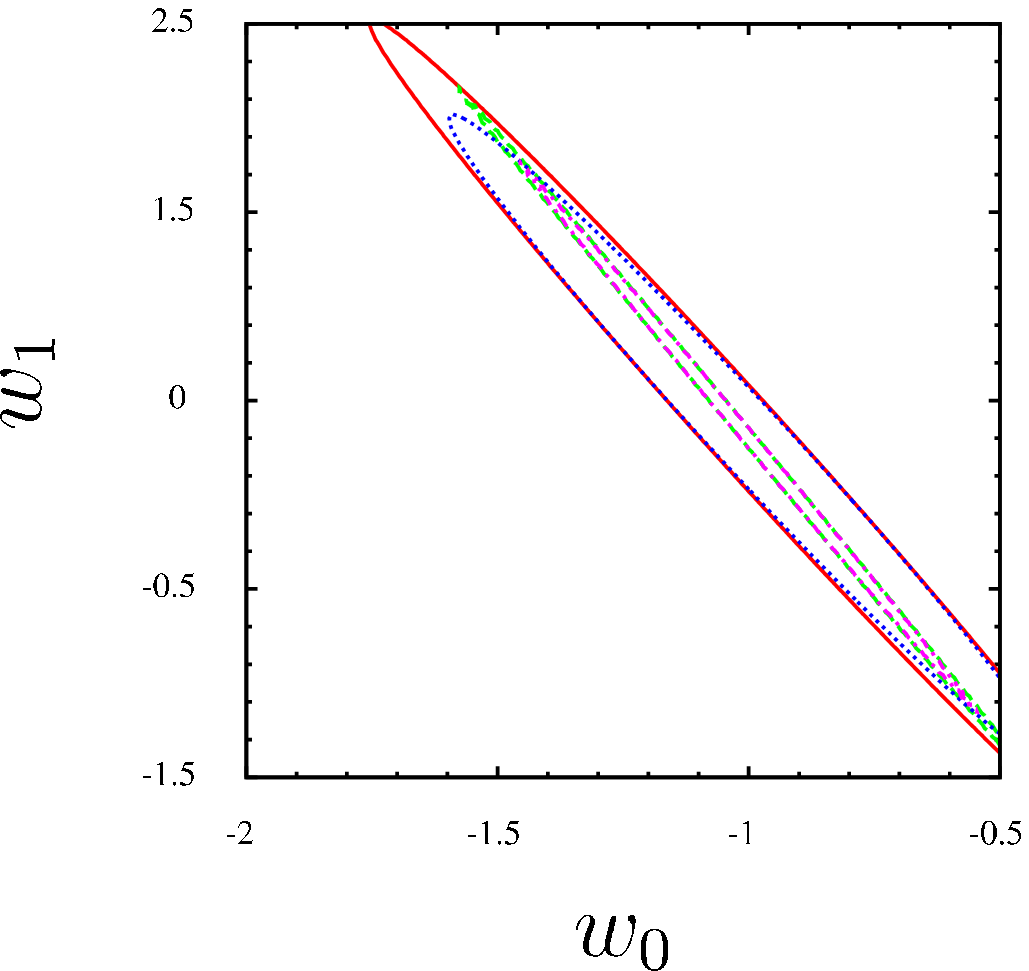}} &\hspace{10mm}
\scalebox{0.7}{\includegraphics{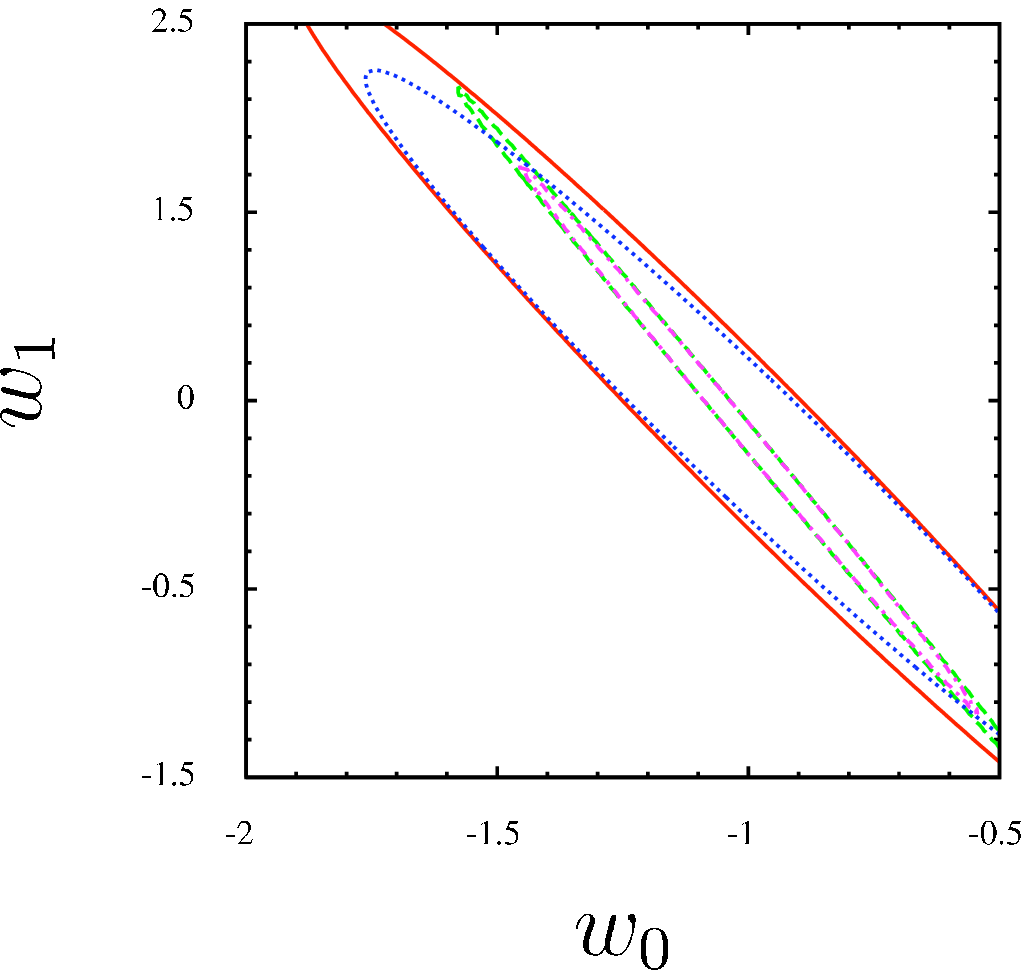}}
\end{tabular}
\caption{Expected 1$\sigma$ contours on the $w_0$ vs. $w_1$ plane.  On
the left panel, we fixed the neutrino masses to the fiducial value.  On
the right panel, we marginalized over the neutrino masses. Contours
which correspond to the attainable future constraints from CMB only (red
solid line), CMB$+$cross correlation (blue dotted line), CMB$+$LSS
(green dashed line) and all combined (purple dash-dotted line) are
shown.}
\label{fig:w0_w1}
\end{center}
\end{figure}

Next we discuss the expected constraints on the dark energy evolutions,
i.e., $w_0$ and $w_1$.  In Fig.~\ref{fig:w0_w1}, we plot the future
constraints on the time evolution of dark energy, fixing the neutrino
masses (left panel) and marginalizing over the neutrino masses (right
panel).  As we can see, when we marginalize the masses of neutrinos, the
uncertainty of $w_0$ and $w_1$ becomes larger, in particular, in the
direction parallel to the $w_0$ axis. In fact, the constraint on $w_1$
is not weakened much compared to that on $w_0$.  As mentioned above,
this is because the masses of neutrino and $w_1$ are not so degenerate,
which means that the time dependence of the equation of state for dark
energy is not strongly correlated with the neutrino masses.  It should
also be noticed that the cross correlation of CMB with LSS can help to
some extent to determine the evolution of dark energy even when we
have uncertainties in the absolute values of the neutrino masses.

\section{Summary} \label{sec:summary}

We discussed the simultaneous determination of the neutrino masses and
the evolution of dark energy from future cosmological observations.
After presenting the degeneracies which exist between the neutrino
masses and dark energy sectors, we considered the possibilities of
breaking the degeneracies using the cross correlation of CMB and LSS. We
studied the cross correlation power spectrum in those models in some
details. We also discussed the suppression of perturbation growth and
the ISW effect in the models.  Then we presented the future constraints
on the neutrino masses and the dark energy equation of state from the
Fisher matrix analysis using the expected data from CMB, LSS and their
cross correlation. First we showed the results on the $\Omega_\nu h^2$
vs. $w_0$ plane for the cases with a constant equation of state and a
time-varying equation of state. We can see that the cross correlation of
CMB with LSS can help to determine the time dependence of the equation
of state to some extent. The constraint on the neutrino masses is not
affected much since there is not strong correlation between the neutrino
masses and $w_1$ which represents the time dependence of dark energy
equation of state in our parameterization.  Then we also discussed the
constraints on the time evolution of equation of state, fixing and
marginalizing the neutrino masses.  When we marginalize the neutrino
masses, the constraints on $w_0$ and $w_1$ are weakened. However, we
showed that the cross correlation observation can be useful to determine
the time evolution of the equation of state even if we do not have much
information on the masses of neutrinos.


{\sl Note added:} Recently the paper \cite{Kiakotou:2007pz} appeared on the arXiv,
which discusses the effect of the neutrino masses on the cross correlation.
Also, Ref.~\cite{Lesgourgues:2007ix} appeared more recently which has some overlaps with our analysis.
Ref.~\cite{Lesgourgues:2007ix} gives a qualitatively similar result to ours that the effects of the neutrino masses
on the cross correlation are relatively small.


\bigskip 
\noindent
{\bf Acknowledgment:} 
We acknowledge the use of CMBFAST \cite{Seljak:1996is} package for our
numerical calculations.  T.T. would like to thank
the Japan Society for Promotion of Science for financial support.

\appendix


\begin{thebibliography}{100}


\bibitem{Cleveland:1998nv}
B.~T.~Cleveland {\it et al.},
Astrophys.\ J.\  {\bf 496}, 505 (1998).

\bibitem{Abdurashitov:2003ew}
J.~N.~Abdurashitov {\it et al.}  [SAGE Collaboration],
Nucl.\ Phys.\ Proc.\ Suppl.\  {\bf 118}, 39 (2003).

\bibitem{Kirsten:2003ev}
T.~A.~Kirsten  [GNO Collaboration],
Nucl.\ Phys.\ Proc.\ Suppl.\  {\bf 118}, 33 (2003).


\bibitem{Fukuda:2001nk}
S.~Fukuda {\it et al.}  [Super-Kamiokande Collaboration],
Phys.\ Rev.\ Lett.\  {\bf 86}, 5656 (2001)
[arXiv:hep-ex/0103033].

\bibitem{Aharmim:2005gt}
B.~Aharmim {\it et al.}  [SNO Collaboration],
arXiv:nucl-ex/0502021.

\bibitem{Araki:2004mb}
T.~Araki {\it et al.}  [KamLAND Collaboration],
Phys.\ Rev.\ Lett.\  {\bf 94}, 081801 (2005)
[arXiv:hep-ex/0406035].


\bibitem{Fukuda:1998mi}
Y.~Fukuda {\it et al.}  [Super-Kamiokande Collaboration],
Phys.\ Rev.\ Lett.\  {\bf 81}, 1562 (1998)
[arXiv:hep-ex/9807003].

\bibitem{Ashie:2005ik}
Y.~Ashie {\it et al.}  [Super-Kamiokande Collaboration],
Phys.\ Rev.\ D {\bf 71}, 112005 (2005)
[arXiv:hep-ex/0501064].


\bibitem{Spergel:2003cb}
  D.~N.~Spergel {\it et al.}  [WMAP Collaboration],
  Astrophys.\ J.\ Suppl.\  {\bf 148}, 175 (2003)
  [arXiv:astro-ph/0302209].

\bibitem{Tegmark:2003ud}
  M.~Tegmark {\it et al.}  [SDSS Collaboration],
  Phys.\ Rev.\ D {\bf 69}, 103501 (2004)
  [arXiv:astro-ph/0310723].

\bibitem{Seljak:2004xh}
  U.~Seljak {\it et al.},
  Phys.\ Rev.\ D {\bf 71}, 103515 (2005)
  [arXiv:astro-ph/0407372].

\bibitem{Ichikawa:2004zi}
  K.~Ichikawa, M.~Fukugita and M.~Kawasaki,
  Phys.\ Rev.\ D {\bf 71}, 043001 (2005)
  [arXiv:astro-ph/0409768].

\bibitem{Spergel:2006hy}
  D.~N.~Spergel {\it et al.}  [WMAP Collaboration],
  Astrophys.\ J.\ Suppl.\  {\bf 170}, 377 (2007)
  [arXiv:astro-ph/0603449].

\bibitem{Tegmark:2006az}
  M.~Tegmark {\it et al.},
  Phys.\ Rev.\  D {\bf 74}, 123507 (2006)
  [arXiv:astro-ph/0608632].

\bibitem{neutrino}
  S.~Hannestad,
  JCAP {\bf 0305}, 004 (2003)
  [arXiv:astro-ph/0303076];
  V.~Barger, D.~Marfatia and A.~Tregre,
  Phys.\ Lett.\ B {\bf 595}, 55 (2004)
  [arXiv:hep-ph/0312065];
  P.~Crotty, J.~Lesgourgues and S.~Pastor,
  Phys.\ Rev.\ D {\bf 69}, 123007 (2004)
  [arXiv:hep-ph/0402049];
  G.~L.~Fogli, E.~Lisi, A.~Marrone, A.~Melchiorri, A.~Palazzo, P.~Serra and J.~Silk,
  Phys.\ Rev.\ D {\bf 70}, 113003 (2004)
  [arXiv:hep-ph/0408045];
  A.~Goobar, S.~Hannestad, E.~Mortsell and H.~Tu,
  JCAP {\bf 0606}, 019 (2006)
  [arXiv:astro-ph/0602155];
  G.~Huetsi,
  Astron.\ Astrophys.\  {\bf 459}, 375 (2006)
  [arXiv:astro-ph/0604129];
  M.~Fukugita, K.~Ichikawa, M.~Kawasaki and O.~Lahav,
  Phys.\ Rev.\  D {\bf 74}, 027302 (2006)
  [arXiv:astro-ph/0605362];
  B.~Feng, J.~Q.~Xia, J.~Yokoyama, X.~Zhang and G.~B.~Zhao,
  JCAP {\bf 0612}, 011 (2006)
  [arXiv:astro-ph/0605742];
  M.~Cirelli and A.~Strumia,
  JCAP {\bf 0612}, 013 (2006)
  [arXiv:astro-ph/0607086];
  S.~Hannestad and G.~G.~Raffelt,
  JCAP {\bf 0611}, 016 (2006)
  [arXiv:astro-ph/0607101];
  G.~L.~Fogli {\it et al.},
  Phys.\ Rev.\  D {\bf 75}, 053001 (2007)
  [arXiv:hep-ph/0608060];
  C.~Zunckel and P.~G.~Ferreira,
  JCAP {\bf 0708}, 004 (2007)
  [arXiv:astro-ph/0610597];
  J.~R.~Kristiansen, H.~K.~Eriksen and O.~Elgaroy,
  Phys.\ Rev.\  D {\bf 74}, 123005 (2006);
  J.~R.~Kristiansen, O.~Elgaroy and H.~Dahle,
  Phys.\ Rev.\  D {\bf 75}, 083510 (2007);

\bibitem{Tonry:2003zg}
J.~L.~Tonry {\it et al.}  [Supernova Search Team Collaboration],
Astrophys.\ J.\  {\bf 594}, 1 (2003)
[arXiv:astro-ph/0305008].

\bibitem{SN}
  P.~Astier {\it et al.}  [The SNLS Collaboration],
  Astron.\ Astrophys.\  {\bf 447}, 31 (2006)
  [arXiv:astro-ph/0510447];
  A.~G.~Riess {\it et al.},
  arXiv:astro-ph/0611572.
  W.~M.~Wood-Vasey {\it et al.},
  arXiv:astro-ph/0701041;
  T.~M.~Davis {\it et al.},
  arXiv:astro-ph/0701510.


\bibitem{MacTavish:2005yk}
  C.~J.~MacTavish {\it et al.},
  Astrophys.\ J.\  {\bf 647}, 799 (2006)
  [arXiv:astro-ph/0507503].

\bibitem{DE_curv}
  K.~Ichikawa and T.~Takahashi,
  Phys.\ Rev.\  D {\bf 73}, 083526 (2006)
  [arXiv:astro-ph/0511821];
  K.~Ichikawa, M.~Kawasaki, T.~Sekiguchi and T.~Takahashi,
  JCAP {\bf 0612}, 005 (2006)
  [arXiv:astro-ph/0605481];
  K.~Ichikawa and T.~Takahashi,
  JCAP {\bf 0702}, 001 (2007)
  [arXiv:astro-ph/0612739];
  K.~Ichikawa and T.~Takahashi,
  arXiv:0710.3995 [astro-ph];
  J.~L.~Crooks, J.~O.~Dunn, P.~H.~Frampton, H.~R.~Norton and T.~Takahashi,
  Astropart.\ Phys.\  {\bf 20}, 361 (2003)
  [arXiv:astro-ph/0305495];
  Y.~g.~Gong and Y.~Z.~Zhang,
  Phys.\ Rev.\  D {\bf 72}, 043518 (2005)
  [arXiv:astro-ph/0502262];
  G.~B.~Zhao, J.~Q.~Xia, H.~Li, C.~Tao, J.~M.~Virey, Z.~H.~Zhu and X.~Zhang,
  Phys.\ Lett.\  B {\bf 648}, 8 (2007)
  [arXiv:astro-ph/0612728];
  Y.~Gong, Q.~Wu and A.~Wang,
  arXiv:0708.1817 [astro-ph];
  D.~Rapetti and S.~W.~Allen,
  arXiv:0710.0440 [astro-ph].




\bibitem{Hannestad:2005gj}
  S.~Hannestad,
  Phys.\ Rev.\ Lett.\  {\bf 95}, 221301 (2005)
  [arXiv:astro-ph/0505551].

\bibitem{Pogosian:2005ez}
  L.~Pogosian, P.~S.~Corasaniti, C.~Stephan-Otto, R.~Crittenden and R.~Nichol,
  Phys.\ Rev.\  D {\bf 72}, 103519 (2005)
  [arXiv:astro-ph/0506396].


\bibitem{Crittenden:1995ak}
R.~G.~Crittenden and N.~Turok,
Phys.\ Rev.\ Lett.\  {\bf 76}, 575 (1996)
[arXiv:astro-ph/9510072].



\bibitem{Peiris:2000kb}
H.~V.~Peiris and D.~N.~Spergel,
Astrophys.\ J.\  {\bf 540}, 605 (2000)
[arXiv:astro-ph/0001393].


\bibitem{Cooray:2001ab}
A.~Cooray,
Phys.\ Rev.\ D {\bf 65}, 103510 (2002)
[arXiv:astro-ph/0112408].


\bibitem{Fosalba:2003iy}
  P.~Fosalba and E.~Gaztanaga,
  Mon.\ Not.\ Roy.\ Astron.\ Soc.\  {\bf 350}, L37 (2004)
  [arXiv:astro-ph/0305468].

\bibitem{Boughn:nature427}
  S.~P.~Boughn and R.~G.~Crittenden,
  Nature {\bf 247}, 45 (2004).

\bibitem{Boughn:2004zm}
  S.~P.~Boughn and R.~G.~Crittenden,
  New Astron.\ Rev.\  {\bf 49}, 75 (2005)
  [arXiv:astro-ph/0404470].

\bibitem{Fosalba:NMRAS}
  P.~Fosalba and E.~Gaztanaga
  Mon.\ Not.\ Roy.\ Astron.\ Soc. {\bf 350}, 37 (2004).

\bibitem{Fosalba:2003ge}
  P.~Fosalba, E.~Gaztanaga and F.~Castander,
  Astrophys.\ J.\  {\bf 597}, L89 (2003)
  [arXiv:astro-ph/0307249].

\bibitem{Scranton:2003in}
  R.~Scranton {\it et al.}  [SDSS Collaboration],
  arXiv:astro-ph/0307335.

\bibitem{Afshordi:2003xu}
  N.~Afshordi, Y.~S.~Loh and M.~A.~Strauss,
  Phys.\ Rev.\ D {\bf 69}, 083524 (2004)
  [arXiv:astro-ph/0308260].

\bibitem{Padmanabhan:2004fy}
  N.~Padmanabhan, C.~M.~Hirata, U.~Seljak, D.~Schlegel, J.~Brinkmann and D.~P.~Schneider,
  Phys.\ Rev.\  D {\bf 72}, 043525 (2005)
  [arXiv:astro-ph/0410360].

\bibitem{Cabre:2006qm}
  A.~Cabre, E.~Gaztanaga, M.~Manera, P.~Fosalba and F.~Castander,
  Mon.\ Not.\ Roy.\ Astron.\ Soc.\ Lett.\  {\bf 372}, L23 (2006)
  [arXiv:astro-ph/0603690].

\bibitem{Giannantonio:2006du}
  T.~Giannantonio {\it et al.},
  Phys.\ Rev.\  D {\bf 74}, 063520 (2006)
  [arXiv:astro-ph/0607572].

\bibitem{McEwen:2006my}
  J.~D.~McEwen, P.~Vielva, M.~P.~Hobson, E.~Martinez-Gonzalez and A.~N.~Lasenby,
  Mon.\ Not.\ Roy.\ Astron.\ Soc.\  {\bf 373}, 1211 (2007)
  [arXiv:astro-ph/0602398].

\bibitem{Garriga:2003nm}
J.~Garriga, L.~Pogosian and T.~Vachaspati,
Phys.\ Rev.\ D {\bf 69}, 063511 (2004)
[arXiv:astro-ph/0311412].


\bibitem{Pogosian:2004wa}
L.~Pogosian,
JCAP {\bf 0504}, 015 (2005)
[arXiv:astro-ph/0409059].




\bibitem{Hu:2004yd}
  W.~Hu and R.~Scranton,
  Phys.\ Rev.\ D {\bf 70}, 123002 (2004)
  [arXiv:astro-ph/0408456].



\bibitem{Corasaniti:2005pq}
  P.~S.~Corasaniti, T.~Giannantonio and A.~Melchiorri,
  Phys.\ Rev.\ D {\bf 71}, 123521 (2005)
  [arXiv:astro-ph/0504115].


\bibitem{Gaztanaga:2004sk}
E.~Gaztanaga, M.~Manera and T.~Multamaki,
arXiv:astro-ph/0407022.



\bibitem{Cooray:2005px}
  A.~Cooray, P.~S.~Corasaniti, T.~Giannantonio and A.~Melchiorri,
  Phys.\ Rev.\ D {\bf 72}, 023514 (2005)
  [arXiv:astro-ph/0504290].


\bibitem{Afshordi:2004kz}
N.~Afshordi,
Phys.\ Rev.\ D {\bf 70}, 083536 (2004)
[arXiv:astro-ph/0401166].

\bibitem{Linder:2002et}
  E.~V.~Linder,
  Phys.\ Rev.\ Lett.\  {\bf 90}, 091301 (2003)
  [arXiv:astro-ph/0208512].


\bibitem{Caldwell:2003vp}
  R.~R.~Caldwell, M.~Doran, C.~M.~Mueller, G.~Schaefer and C.~Wetterich,
  Astrophys.\ J.\  {\bf 591}, L75 (2003)
  [arXiv:astro-ph/0302505].



\bibitem{Seljak:1996is}
  U.~Seljak and M.~Zaldarriaga,
  Astrophys.\ J.\  {\bf 469}, 437 (1996)
  [arXiv:astro-ph/9603033].



\bibitem{Hu:1997vi}
  W.~Hu and D.~J.~Eisenstein,
  Astrophys.\ J.\  {\bf 498}, 497 (1998)
  [arXiv:astro-ph/9710216].


\bibitem{Bean:2003fb}
  R.~Bean and O.~Dore,
  Phys.\ Rev.\ D {\bf 69}, 083503 (2004)
  [arXiv:astro-ph/0307100].



\bibitem{Tegmark:1996bz}
  M.~Tegmark, A.~Taylor and A.~Heavens,
  Astrophys.\ J.\  {\bf 480}, 22 (1997)
  [arXiv:astro-ph/9603021].




\bibitem{Jungman:1995av}
  G.~Jungman, M.~Kamionkowski, A.~Kosowsky and D.~N.~Spergel,
  Phys.\ Rev.\ Lett.\  {\bf 76}, 1007 (1996)
  [arXiv:astro-ph/9507080].


\bibitem{Tegmark:1997rp}
  M.~Tegmark,
  Phys.\ Rev.\ Lett.\  {\bf 79}, 3806 (1997)
  [arXiv:astro-ph/9706198].


\bibitem{Eisenstein:1998hr}
  D.~J.~Eisenstein, W.~Hu and M.~Tegmark,
  Astrophys.\ J.\  {\bf 518}, 2 (1998)
  [arXiv:astro-ph/9807130].


\bibitem{planck}
{\tt http://www.rssd.esa.int/Planck}

\bibitem{LSST}
{\tt http://www.lsst.org/}

\bibitem{Lesgourgues:2004ps}
  J.~Lesgourgues, S.~Pastor and L.~Perotto,
  Phys.\ Rev.\ D {\bf 70}, 045016 (2004)
  [arXiv:hep-ph/0403296].



\bibitem{Kiakotou:2007pz}
  A.~Kiakotou, O.~Elgaroy and O.~Lahav,
  arXiv:0709.0253 [astro-ph].

\bibitem{Lesgourgues:2007ix}
  J.~Lesgourgues, W.~Valkenburg and E.~Gaztanaga,
  arXiv:0710.5525 [astro-ph].

\end{thebibliography}
\end{document}